\documentstyle[preprint,eqsecnum,aps]{revtex}
\newcommand{\beq}{\begin{equation}}
\newcommand{\eeq}{\end{equation}}

\def\real{\rm I\! R}
\def\sRR{{\sl \hbox{I\kern-.2em\hbox{R}}}}
\def\CC{\hbox{C\kern -.58em {\raise .4ex \hbox{$\scriptscriptstyle |$}}
 \kern-.55em {\raise .53ex \hbox{$\scriptscriptstyle |$}} }}
\def\ZZ{{{\rm Z}\kern-.28em{\rm Z}}}
\def\lra{\vbox{\baselineskip 2pt\hbox
 {$\leftrightarrow$}\par\hbox{$\nabla^a$}}}
\begin{document}
\draft
\preprint{WISC-MILW-94-TH-23}
\title{Quantum field theory in Lorentzian universes-from-nothing}
\author{John L. Friedman}
\address{
Department of Physics, University of Wisconsin\\
Milwaukee, Wisconsin 53201}
\author{Atsushi Higuchi}
\address{Institute for Theoretical Physics\\
     University of Bern\\
     Sidlerstrasse 5, CH-3012 Bern\\
     Switzerland}
\maketitle
\begin{abstract}

We examine quantum field theory in spacetimes that are time
nonorientable but have no other causal pathology.  These are Lorentzian
universes-from-nothing, spacetimes with a single spacelike boundary
that nevertheless have a smooth Lorentzian metric. A
time-nonorientable, spacelike hypersurface serves as a generalized
Cauchy surface, a surface on which freely specified initial data
for wave equations have unique global time evolutions.  A simple
example is antipodally identified deSitter space.  Classically, such
spacetimes are locally indistinguishable from their globally hyperbolic
covering spaces.

   The construction of a quantum field theory is more problematic.  Time
nonorientability precludes the existence of a global algebra of
observables, and hence of global states, regarded as positive linear
functions on a global algebra.  One can, however, define a family of
local algebras on an atlas of globally hyperbolic subspacetimes, with
overlap conditions on the intersections of neighborhoods.  This family
locally coincides with the family of algebras on a globally
hyperbolic spacetime; and one can ask whether a sensible quantum field
theory is obtained if one defines a state as an assignment of a
positive linear function to every local algebra.  We show, however,
that the extension of a generic positive linear function from a single
algebra to the collection of all local algebras violates positivity: one
cannot find a collection of quantum states satisfying the physically
appropriate overlap conditions.

One can overcome this difficulty by artificially restricting the size
of neighborhoods in a way that has no classical counterpart.
Neighborhoods in the atlas must be small enough that the union of any
pair is time-orientable.  Correlations between field operators at a
pair of points are then defined only if a curve joining the points lies
in a single neighborhood.  Any state on one neighborhood of an
atlas can be extended to a collection of states on the atlas, and the
structure of local algebras and states is thus locally
indistinguishable from quantum field theory on a globally hyperbolic
spacetime.  But the artificiality of the size restriction on
neighborhoods means that the structure is not a satisfactory global
field theory.  The structure is not unique, because there is no unique
maximal atlas.  The resulting theory allows less information than
quantum field theory in a globally hyperbolic spacetime, because there
are always sets of points in the spacetime for which no correlation
function is defined.  Finally, in showing that one can extend a local
state to a collection of states, we use an antipodally symmetric state
on the covering space, a state that would not yield a sensible state on
the spacetime if all correlations could be measured.
\end{abstract}
\pacs{PACS: 04.20.Gz, 04.60.+v, 04.20.Cv}
\narrowtext
\section{Introduction}
\label{sec:intro}
In a Lorentzian path-integral approach to quantum gravity, one can, as
in the Euclidean case, imagine constructing a wave function of the
universe from a sum over all Lorentzian 4-geometries with a single
spacelike boundary.  Spacetimes of this kind provide the only examples
of topology change in which one can have a smooth, nondegenerate
Lorentz metric without closed timelike curves; instead, the spacetimes
are time nonorientable.

The simplest examples of such spacetimes have
the topology of a finite timelike cylinder, $S^3\times \real^{+}$,
with diametrically opposite points of its past spherical boundary
identified.  This is the topology of antipodally identified deSitter
space.  It is a 4-dimensional analog of the M\"obius strip, which can
be constructed from a finite 2-dimensional timelike cylinder by
identifying diametrically opposite points of its circular past
boundary, $\tilde \Sigma$ (see Fig.~\ref{f1}).  A more familiar
representation of the same strip is shown in Fig.~\ref{f2}, whose
median circle $\Sigma$ was the one just constructed by identifying
points of $\tilde \Sigma$. The orientable double-covering space of the
strip is a cylinder $\tilde  M$ of double the timelike length
(Fig.~\ref{f2}), and $M$ is constructed from $\tilde  M$ by identifying
antipodal points.  If the covering space has the metric of a flat,
timelike cylinder, the M\"obius strip $M$ will be time nonorientable
with a locally flat metric and a timelike Killing vector (defined
globally only up to sign) perpendicular to $\Sigma$.  If the covering
spacetime $\tilde  M$ is given the 2-dimensional deSitter metric, the
M\"obius strip will acquire the metric of antipodally identified
deSitter space.

As recent authors have noted, the time-nonorientability of these
spacetimes prevents one from carrying through the standard construction
of a Fock space or a Weyl algebra of observables \cite{CTw,CTh,sw}.
Kay \cite{CTh} requires existence of a globally
defined *-algebra and imposes what he terms the ``F-locality
condition'', which demands
in essence, that the *-algebra satisfy the canonical commutation
relations in a neighborhood of any point with respect to one time
orientation.  Under these conditions he proves that the spacetime must
be time orientable.  An independent study by Gibbons \cite{CTw}
concludes that one is forced to use a real (i.e., noncomplex) Hilbert
space to describe quantum field theory in some time-nonorientable
spacetimes including antipodally identified deSitter space.  This also
suggests that one cannot construct a globally defined conventional
quantum field theory in a time-nonorientable spacetime.

The conclusion is surprising, because an observer in antipodally identified
deSitter space, $M$, cannot classically distinguish the spacetime from
deSitter space, $\tilde  M$. The past of a timelike worldline in $M$
(defined by a choice of orientation near the worldline) is isometric to
the past of either of the two corresponding worldlines in $\tilde  M$.
In fact, the Cauchy problem is well-defined on $M$ for fields with
initial data on $\Sigma$, \footnote{This is because the lift to
$\tilde \Sigma$ of data on $\Sigma$ will evolve to an antipodally
invariant field on $\tilde  M$.  The field on $\tilde  M$ will therefore
be the lift of a solution to the field equation on $M$.} and the
solutions will be identical to those seen by an observer on
$\tilde \Sigma$ who travels along the corresponding worldline and sees
the same data.

We are concerned in this paper with whether one can evade these global
results by piecing together local quantum algebras and states.  We find
that one {\em can} evade Kay's no-go theorem if, instead of a
globally-defined $^*$-algebra one demands only a set of
$^*$-algebras, each defined in a local neighborhood.  Overlap
conditions on the $^*$-algebras then ensure that the local algebraic
structure coincides with that on a globally hyperbolic spacetime.  One
would like to use this structure of local algebras to define quantum
states as collections of positive linear functions (plf's) on the local
algebras, again with consistency conditions on the intersection of
globally hyperbolic neighborhoods.  We find, however, that if one
considers the algebras of observables on an atlas consisting of {\em
all} globally hyperbolic subspacetimes that inherit their causal
structure from the spacetime $M, g$, then one cannot consistently
define states.  In particular, if the union of a pair of neighborhoods
is time nonorientable, one cannot consistently extend a generic plf to
the pair of neighborhoods without violating positivity.

One can define a collection of local states on {\em smaller} atlases,
restricted so that the union of any two neighborhoods is time
orientable.  The  collection of algebras and states is then locally
indistinguishable from that on the globally hyperbolic covering space
of any Lorentzian universe-from-nothing.  In particular, any local
state can be consistently extended to a collection of states on all algebras
associated with the atlas.
But the restriction on the size of
neighborhoods amounts to a restriction on the size of regions over
which one can define correlations between field operators, and this has
unpleasant implications.  The specification of a collection of states
on the neighborhoods that cover and share an initial value surface does
not {\em uniquely} determine a time-evolution: The extension to a
collection of states on the set of all algebras is not unique.
In addition, in showing that one can extend a local
state to a collection of states, we use an antipodally symmetric state
on the covering space, and such a state would not yield a
well-defined state on the spacetime if all correlations could be measured.
Finally, the families of states and algebras depend on the choice of
atlas, and there is no unique maximal atlas.

The ability to construct a family of states and algebras that agrees locally
with that of a globally hyperbolic spacetime relies on the fact that
the spacetimes we consider, although time nonorientable, have no closed
timelike curves (CTCs).  The simplest example above, the flat M\"obius
strip, has CTCs if one extends the strip to a timelike thickness
greater than its circumference.  (These are smooth timelike curves
$c(\lambda)$ that intersect the same point twice; the tangent vectors at
the point of intersection have opposite time-orientation.)
In the nonchronal region (the region with CTCs) nearby points that are
spacelike separated with respect to the causal structure of a globally
hyperbolic neighborhood are joined by timelike curves in the full
spacetime.  Thus events that are locally spacelike-separated will
influence one another, and one naturally expects that field operators
at points
whose local separation is spacelike will fail to commute or that there
will be restrictions on algebraic states in the local neighborhoods.
In Yurtsever's \cite{CFo}
generalization of the algebraic approach to quantum field theory to
spacetimes with CTCs, massive scalar field theory will not
in general have local algebras of observables that agree with the ordinary
local algebras of observables associated (by the usual construction) with
sufficiently small globally hyperbolic neighborhoods of each spacetime point.
On the other hand, in Kay's approach \cite{CTh}, one {\em requires}
such an agreement
(F-quantum compatibility).  This requirement can be implemented both for
massless and massive scalar field theories
in some spacetimes with CTCs \cite{CTh,FH} though it is
not clear how these works can be extended to more general spacetimes.
Antipodally-identified deSitter space avoids CTCs by expanding
rapidly enough that timelike curves that loop through $\Sigma$ cannot
quite return to their starting point.

\begin{figure}
\caption{The double cover $\tilde M$ is related to $M$ by the identification
of antipodal points, $A$ and $A'$.} \label{f3}
\end{figure}
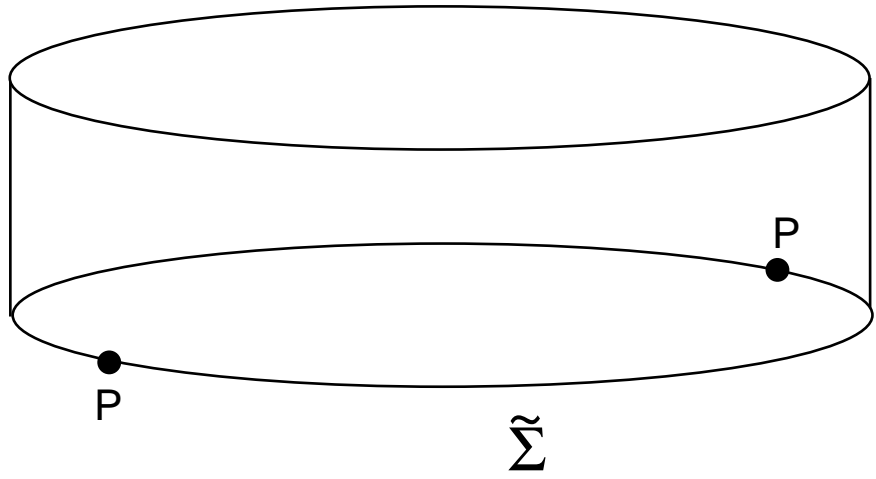
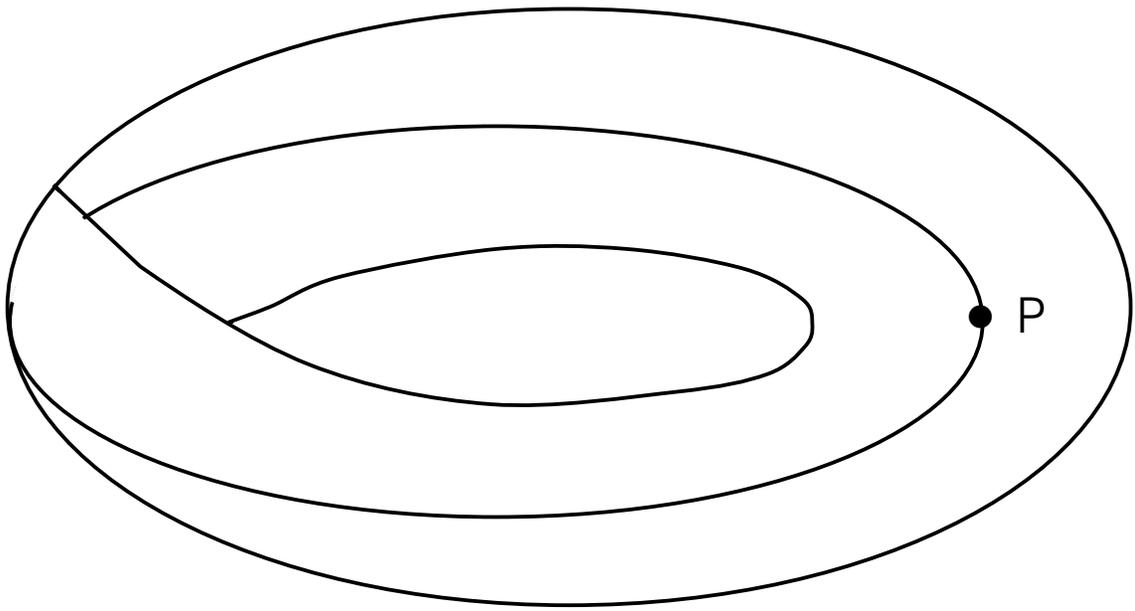
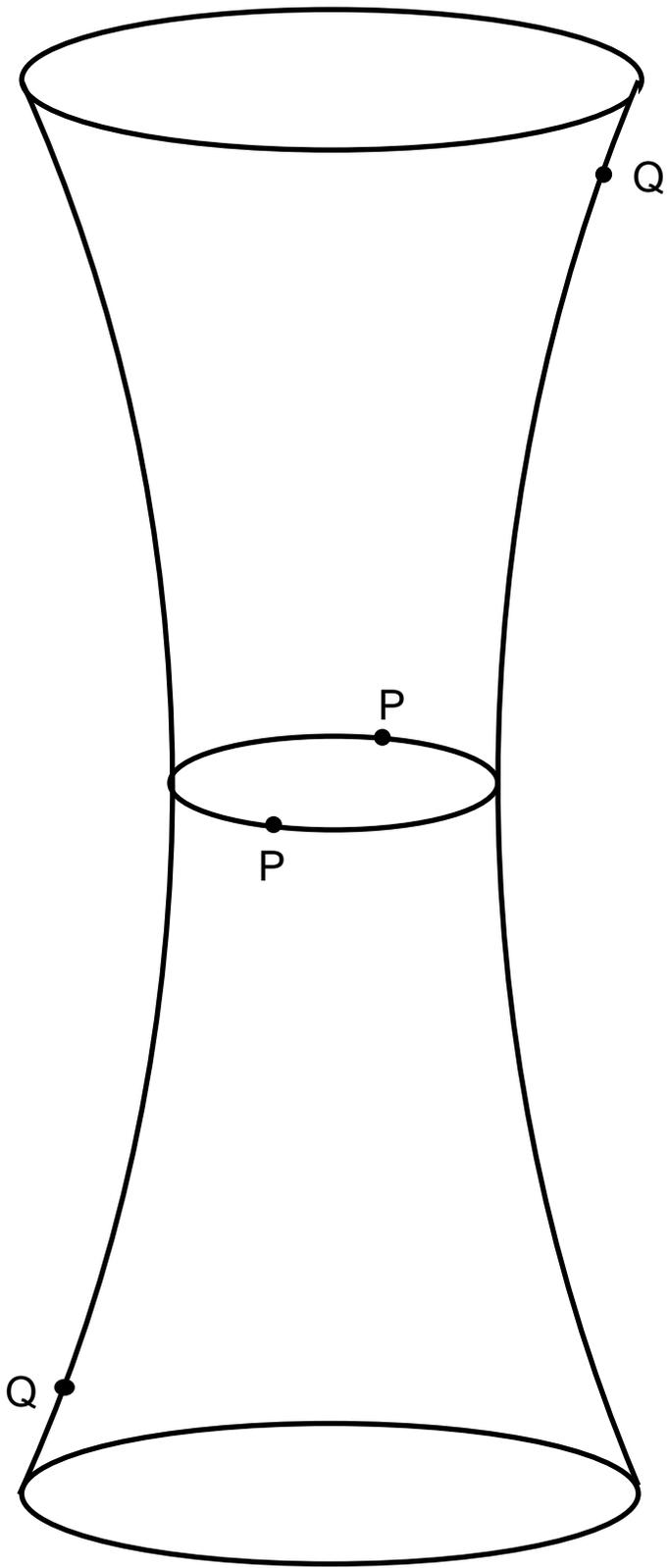

Lorentzian universes with no past boundary and no CTCs can be
constructed in the way shown in Fig.~\ref{f3} from any compact
orientable 3-manifold $\tilde\Sigma$ that admits an involution -- a
diffeomorphism $I$ that acts freely on $\tilde\Sigma$ and for which
$I^2 = 1$.  That is, if $T:\sRR\rightarrow \sRR$ is the map
$t\mapsto -t$, the manifold is the quotient
\beq
M = (\sRR\times\tilde\Sigma)/(T\times I).
\label{lor}\eeq
Countably many 3-manifolds admit free involutions, including  all lens
spaces and most other spherical spaces; each gives rise to a
topologically distict class of Lorentzian universes-from-nothing.

\section{Quantum field theory without a choice of time-orientation}
\label{sec:notime}

\subsection{Minkowski space}

We begin with a Fock-space framework for concreteness and to make the
subsequent algebraic treatment of oppositely oriented observers more
transparent.

The quantum theory of a neutral scalar field on Minkowski space can be
described in terms of the space $V$ of real solutions to the
Klein-Gordon
equation,
\begin{equation}
Kf := (-\nabla_a \nabla^a + m^2)f = 0\,, \label{set}
\end{equation}
which are finite in the norm,
\begin{equation}
\int_{\Sigma} dS\,(\vert n^a\nabla_a f\vert^2 + \kappa^2 f^2)\,, \label{eit}
\end{equation}
where $\Sigma$ is $t=$constant surface,  $n_a$ is a
unit normal to $\Sigma$ and $\kappa^2$ is an arbitrary positive constant.
One makes $V$ into a complex vector space by
a choice of complex structure $J$, which in turn relies on choosing an
orientation of time.  If ${\cal O}$ is the orientation for which $n^a$ is
future-pointing, then
\begin{equation}
Jf = i(f^{(+)} - f^{(-)})\,, \label{cstruct}
\end{equation}
where $f^{(+)}$ is the positive frequency part of $f$ with respect to
${\cal O}$.  A reversal of time-orientation reverses the assignment of
positive and negative frequencies. By itself, however, the space of
real solutions is independent of orientation.

Given a choice ${\cal O}$ of time-orientation, one can define a symplectic
product $\omega$ on $V$ by writing
\begin{equation}
\omega(f,g) = \int_{\Sigma}dS_a f\lra g\,,
\label{nit}
\end{equation}
where
\begin{equation}
dS_a = n_a dS \label{twy}
\end{equation}
with $n^a$ the normal that is future directed with respect to ${\cal O}$.
The corresponding inner product on $V$ has the form
\begin{equation}
\langle f\vert g \rangle = \frac{1}{2}\,\omega(f,Jg)
+ \frac{i}{2}\,\omega(f,g)\,. \label{twtw}
\end{equation}
The completion of $V$ in the inner product
$\scriptstyle{\langle\ \vert\ \rangle}$ with
complex structure $J$ is the 1-particle Hilbert space ${\cal H}$ of
the free scalar field, and the corresponding Fock space is
\begin{equation}
{\cal F} = \CC + {\cal H} + {\cal H}\otimes_s{\cal H} + \cdots.  \label{twth}
\end{equation}

An observer with time-orientation $\check{\cal O}$
opposite to ${\cal O}$ will
use normal
\begin{equation}
\check n_a = -n_a\,, \label{twfi}
\end{equation}
surface element
\begin{equation}
d\check S_a = \check n_a dS = -dS_a\,, \label{twsi}
\end{equation}
and symplectic structure
\begin{equation}
\check\omega(f,g) = \int_{\Sigma}d\check S_a
f\lra g = -\omega(f,g)\,.  \label{twse}
\end{equation}
The complex structure on $V$ similarly changes sign, because the positive
and negative frequency parts of $f\in V$ are interchanged:
\begin{equation}
\check J = -J\,. \label{twei}
\end{equation}
Eq.~(\ref{twtw}) then implies that oppositely oriented observers assign to
the same pair of real solutions (and hence to the same 1-particle state)
complex conjugate inner products,
\begin{eqnarray}
\check\langle f\vert g\check\rangle & = & \frac{1}{2}\,
\check\omega(f,\check J g) + \frac{i}{2}\,\check\omega(f,g) \nonumber \\
 & = & \frac{1}{2}\,\omega(f,Jg) - \frac{i}{2}\,\omega(f,g) \nonumber \\
 & = & \overline{\langle f\vert g\rangle}\,. \label{twni}
\end{eqnarray}
The map $|\rangle \mapsto |\check\rangle$ induces an antiunitary map
${\cal I}:{\cal F}\longrightarrow \check{\cal F}$, with
\begin{equation}
{\cal I}  \alpha\vert
f_1\rangle\otimes_s\cdots\otimes_s\vert f_k\rangle
=\bar{\alpha}\vert f_1\check\rangle\otimes_s\cdots\otimes_s\vert
f_k\check\rangle\,.
\end{equation}
A pure state $[\Psi]$ can be regarded as assigning to time-orientations
${\cal O}$ and $\check{\cal O}$ vectors $\Psi\in{\cal F}$ and
$\check\Psi\in\check{\cal F}$, with $\check\Psi={\cal I}\Psi$;
more generally, an  (algebraic) state $[\rho]$
assigns states $\rho$ and
$\check\rho$ to orientations ${\cal O}$ and $\check{\cal O}$.

The Heisenberg field operator $[\hat{\phi}]$ similarly assigns to
time-orientations ${\cal O}$ and $\check{\cal O}$ operators
$\hat{\phi}$ and $\check\phi $, acting on ${\cal F}$ and $\check  {\cal
F}$ respectively. For orientation ${\cal O}$, smeared field operators
\begin{equation}
\hat{\phi}(F) = \int \hat\phi(x) F(x) d^4x \,,
\label{thfo}
\end{equation}
have commutation relations
\begin{equation}
[ \hat{\phi}(E),\hat{\phi}(F) ]
= i\int E(x)(G_{\rm adv}-G_{\rm ret})(x,y)F(y) d^4x d^4y\,,
\label{thfi}
\end{equation}
where $G_{\rm adv}$ ($G_{\rm ret}$) is the advanced (retarded) Green
function.
An observer with opposite time-orientation $\check{\cal O}$ will adopt
the opposite sign for the commutator, because she will use the opposite
definitions $\check G _{\rm adv}= G_{\rm ret}$ and $\check G _{\rm ret}=
G_{\rm adv}$:
\begin{equation}
[ \check\phi (E),\check\phi (F) ] = i\int E(x)(\check G _{\rm
adv} -\check G _{\rm ret})(x,y)F(y) d^4x d^4y\,.
\label{thsi}
\end{equation}

The structure of the algebra is clearer if one uses the fact that
each smeared field operator can be written as the symplectic product of
$\hat\phi$ with a real solution $f$ to the Klein-Gordon equation:
\begin{equation}
\hat\phi(F) = \omega(\hat{\phi},f) = \int_{\sigma}
dS\, [\hat{\phi}(x) n_a\nabla^a f(x) -\hat{\pi}(x)f(x) ]\,,
\label{phif}
\end{equation}
where
\begin{equation}
f(x) = \int (G_{\rm adv}-G_{\rm ret})(x,y)F(y) d^4x\,.
\label{fx}\end{equation}
The canonical commutation relations are simply
\begin{equation}
[ \omega(\hat{\phi},f),\omega(\hat{\phi},g) ] = i \omega(f,g)\,.
\label{ccr3}
\end{equation}

Expectation values of elements in the algebra depend on orientation in
the manner,
\begin{equation}
\check\langle \check\Psi\vert  \check\phi (F)\cdots\check\pi (G)
\vert\check\Psi\check\rangle
= \overline{\langle\Psi\vert \hat\phi(F)\cdots (-\hat\pi(G))
\vert\Psi\rangle}\,, \label{tpsi}
\end{equation}
or, for general algebraic state
$[\rho]$,
\begin{equation}
\check\rho  [ \check\phi (F)\cdots\check\pi (G) ]
= \overline{\rho [\hat\phi(F)\cdots (-\hat\pi(G)) ]}\,. \label{trho0}
\end{equation}

Note that Eq.~(\ref{trho0}) follows from the relation
between $\rho$ and $\check\rho$ acting on a string of smeared $\check\phi $'s,
\begin{equation}
\check\rho  [ \check\phi (F)\cdots\check\phi (G) ]
= \overline{\rho [\hat\phi(F)\cdots \hat\phi(G) ]}\,,
\end{equation}
because $\hat\pi(F)=-\hat\phi(\partial_t F)$
and $\check\pi (F)=-\check\phi (\partial_{\check  t} F)$,
where $\check  t = -t$.

\subsection {Globally hyperbolic spacetimes}
We now generalize the treatment of a scalar field to arbitrary globally
hyperbolic spacetimes $M, g$.  We use an algebraic approach for quantum
fields (see Haag \cite{CSet} for a review) developed for curved
spacetimes by a number of earlier authors (Ashtekar and Magnon
\cite{CSe}, Isham \cite{CNi}, Kay \cite{CEi}, H\'aj\'{\i}\v{c}ek
\cite{CTe}, Dimock \cite{CEl}, Fredenhagen and Haag \cite{CSit}, and
Kay and Wald \cite{CTht,CTwe}).

Corresponding to orientations ${\cal O}$ and $\check{\cal O}$, one
defines abstract algebras ${\cal A}$ and $\check {\cal A}$ as the free
complex algebras generated by symbols of the form $\{\hat\phi(F),
F\in C^\infty_0(M)\}$ and $\{\check\phi (F), F\in C^\infty_0(M)\}$,
modulo the commutation relations,
\begin{eqnarray}
\left[ \hat{\phi}(E),\hat{\phi}(F) \right]
&=& i\int E(x)(G_{\rm adv}-G_{\rm ret})(x,y)F(y) d^4V_x d^4V_y, \nonumber \\
\left[\check\phi (E),\check\phi (F)\right] & =  &
i\int E(x)(\check G _{\rm adv}-\check G _{\rm ret})(x,y)F(y) d^4V_x
d^4V_y\,.
\label{ccr}
\end{eqnarray}

To each globally hyperbolic subspacetime $U, g|_U$ of $M,g$
and each choice of orientation ${\cal O}$ on $U$ corresponds a local
algebra ${\cal A}_{(U,{\cal O})}$ defined as above, with $M$ replaced
by $U$.  On each overlap, $U\bigcap U'$, there is a linear or antilinear
isomorphism ${\cal I}$ between the restrictions of ${\cal A}_{(U,{\cal
O})}$ and ${\cal A}_{(U',{\cal O}')}$ to $\phi$'s smeared with
functions having support on $U\bigcap U'$.  For $F\in C_0^\infty(U\cap
U')$
\begin{eqnarray}
{\cal I}\hat\phi(F) & =  & \hat\phi'(F),\ {\cal I} i = i, \;\;
{\rm if} \;\; {\cal O} = {\cal O}', \nonumber \\
{\cal I}\hat\phi(F) & =  & \check{\phi}'(F),\ {\cal I} i = - i,\;\;
{\rm if}\;\;
{\cal O} \neq {\cal O}'\,.
\label{algiso} \end{eqnarray}
In particular, for $U'=M$ with agreeing orientations, the map ${\cal
I}$ embeds ${\cal A}_{(U,{\cal O})}$ in ${\cal A}_{(M,{\cal O})}$
as a subalgebra.

In the algebraic approach, with a fixed time-orientation, a physical
state is a positive linear function (plf)
$\rho$ on the algebra of
observables.  When the algebra is represented by a set of linear
operators on a Hilbert space, a state $\rho$ is represented by a
density matrix.  Again, one can democratically define a state $[\rho]$ as an
assignment of a
plf, $\rho$ or
$\check\rho $, to each choice of orientation (${\cal O}$ or
$\check{\cal O}$), where $\check\rho \in \check{\cal F}
\otimes\check{\cal F}^*$.  Formally, a state $[\rho]$ is an
equivalence class of pairs, $(\rho,{\cal O})$, satisfying,
$$
(\rho',{\cal O}') \equiv (\rho,{\cal O}) \Longleftrightarrow
$$
\begin{equation}
\rho' = \rho\;\; {\rm if}\;\; {\cal O}' = {\cal O}\;\;
{\rm and}\;\;
\rho' = \check\rho \;\; {\rm if}\;\;{\cal O}' \neq \check{\cal O}\,,
\label{foth}
\end{equation}
where, for an arbitrary string of operators,
\begin{equation}
\check\rho  [ \check\phi (F)\cdots\check\phi (G) ]
= \overline{\rho [\hat\phi(F)\cdots \hat\phi(G) ]}\,,
\label{trho1}
\end{equation}
implying
\begin{equation}
\check\rho  [ \check\phi (F)\cdots\check\pi(G) ]
= \overline{\rho [\hat\phi(F)\cdots (-\hat\pi(G)) ]}\,. \label{trho2}
\end{equation}

The restriction of a state $[\rho]$ to the pair of subalgebras
associated with a globally hyperbolic neighborhood $U$ is a state
$[\rho_U]$ on $U$.  As in Eq.~(\ref{trho0}), states on overlapping
neighborhoods $U$ and $U'$ are related by
\begin{eqnarray} \rho_{(U,{\cal O})}(A) = \rho_{(U',{\cal O}')}(A')\;\;
{\rm if} \;\; {\cal O} = {\cal O}', \nonumber \\
\rho_{(U,{\cal O})}(A) = \overline{\rho_{(U',{\cal O}')}(A')}
= \rho_{(U',{\cal O}')}(A'^\dagger),\;\; {\rm if}\;\; {\cal O} \neq {\cal O}'
\,,
\label{rhocond}
\end{eqnarray}
where $A' = {\cal I} A $, with ${\cal I}$ given by Eq.~(\ref{algiso}).

\section{Quantum field theory on time-nonorientable spacetimes}
\label{sec:noor}
\subsection{Existence of a family of local algebras}

As in the previous section, we will define local algebras of
observables in a neighborhood of each point, show that the algebras on
overlapping regions are isomorphic, and define global states as
positive linear functions on each local algebra that respect the
overlap isomorphisms. The construction uses the fact that there is a
well-defined initial value problem on the spacetimes $M, g$ that we
consider and that one can find an atlas of globally hyperbolic
neighborhoods which inherit their causal structure from $M$.  \\ {\sl
Definition}.  A spacelike hypersurface $\Sigma$ of a spacetime $M, g$
is an {\sl initial value surface} (for the Klein-Gordon equation) if,
for any smooth choice of $\phi$ and its normal derivative on $\Sigma$,
there is a unique $\phi$ on $M$ satisfying $K\phi = 0$. \\ We are
concerned with time-nonorientable spacetimes $M,g$ which have no closed
timelike curves and whose double-covering $\tilde  M, \tilde  g$ is
globally hyperbolic; as noted in Sect.~\ref{sec:intro}, any
hypersurface $\Sigma\subset M$ is an initial value surface if its lift
to $\check M$ is a Cauchy surface.

A local algebra of observables can be defined on any neighborhood
$U\subset M$ for which\\
(i) $U, g|_U$, regarded as a spacetime, is globally
hyperbolic, and \\
(ii) $U$ is connected and {\sl causally convex} \cite{penrose}.\\
An open set $U$ is causally convex if no causal curve in $M$ intersects
$U$ in a disconnected set.  If $U$ is not causally convex, then some
points that are spacelike separated in the spacetime $U, g|_U$
are joined by a null or timelike curve in $M$, and the commutation
relations for field operators can not be deduced by the causal
structure of $U$.  A causally convex neighborhood inherits its causal
structure from $M$.\footnote{Although conditions (i) and (ii) above
resemble what is called the local causality property\cite{penrose}, the
latter is much more restrictive:  the closure of a local causality
neighborhood is required to lie in a geodesically convex normal
neighborhood.}

Let $C=\{ U \}$ be an atlas for $M$, a collection of open sets that cover
$M$, for which\\
(i) and (ii) above hold for each set $U\in C$;\\
(iii) each $U\in C$ has a Cauchy surface that can be completed to an initial
value surface of $M$.\\
One would like to define a collection of local algebras and states on
all oriented subspacetimes $U, g|_U$, satisfying (i)-(iii), where
algebras on overlapping neighborhoods are related by linear or
antilinear isomorphisms ${\cal I}$ and states on overlapping
neighborhoods are related by Eq.~(\ref{rhocond}).  Although a
consistent definition of a collection of states will require an
additional unwanted restriction on the size of neighborhoods, the
structure of local algebras coincides with that of the globally
hyperbolic spacetime.

Let ${\cal C} = \{ (U, {\cal O}), (U,\check{\cal O}),\ldots \}$ be the
collection of all pairs with $U\in C$ and ${\cal O}$ a choice of
orientation for $U$.  Given a neighborhood $U\in C$, we associate with
orientations ${\cal O}$ and $\check{\cal O}$ algebras of observables,
${\cal A}_{(U,{\cal O})}$ and ${\cal A}_{(U,\check{\cal O})}$, using
the fact that $U, g|_U$ is globally hyperbolic to construct the Green
functions $G_{\rm ret}$, $G_{\rm adv}$, $\check G _{\rm ret}=G_{\rm
adv}$, $\check G _{\rm adv}=G_{\rm ret}$.
 \\ {\sl Definition}. The algebras ${\cal
A}_{(U,{\cal O})}$ and ${\cal A}_{(U,\check{\cal O})}$ are the free
complex algebras generated by $\{\hat\phi(F), F\in C_0^\infty(U)\}$ and
$\{\check\phi (F), F\in C_0^\infty(U)\}$ modulo the canonical
commutation relations, (\ref{ccr}). \\
The algebras are related by an antilinear isomorphism, ${\cal I}:{\cal
A}_{(U,{\cal O})}\longrightarrow{\cal A}_{(U,\check{\cal O})}$, given
by
\begin{eqnarray}
{\cal I}\hat\phi(F)&=&\check\phi (F), \nonumber\\
{\cal I} i &=& - i\,. \label{aiso}
\end{eqnarray}
Writing $\hat\pi(F):=-\hat\phi(\nabla_a(n^a F))$,
with $n^a$ the future pointing
normal with respect to orientation ${\cal O}$, we have
\begin{equation}
{\cal I}\hat\pi(F)=-\check\pi (F)\,.
\end{equation}
We thus have a collection of pairs $({\cal A}_U,{\cal O})$, related on
each overlap $U\bigcap U'$ by the linear or antilinear isomorphism given in
Eq.~(\ref{algiso}).
Thus one can consistently define a pair of oppositely oriented algebras
for every globally hyperbolic neighborhood $U$ that inherits its causal
structure from $M$.  By allowing a pair of algebras at each point we
evade Kay's ``F-locality'' condition\cite{CTh}.

\subsection{Nonexistence of states on the family of all local algebras}

Suppose now that one tries to define a state $\rho$ as an assignment of
a plf $\rho_{(U,{\cal O})}$ to the algebra ${\cal A}_{(U,{\cal O})}$ of
each oriented neighborhood $(U,{\cal O})$, satisfying overlap
conditions on intersections of neighborhoods.  That is, in order that
the state $\rho$ look locally like a state on a globally hyperbolic
spacetime, defined in Sect.~\ref{sec:notime}, it must obey the same
conditions on intersections of neighborhoods: For any $A$ in the
subalgebra generated by $\hat\phi(F)\in{\cal A}_U$ with $F\in
C_0^\infty(U\bigcap V)$,
\begin{eqnarray}
\rho_{(U,{\cal O})}(A) &=& \rho_{(U',{\cal O}')}(A')\;\; {\rm if}
\;\; {\cal O} = {\cal O}', \nonumber \\
\rho_{(U,{\cal O})}(A) &=& \overline{\rho_{(U',{\cal O}')}(A')} =
\rho_{(U',{\cal O}')}(A'^\dagger),\;\; {\rm if}\;\; {\cal O} \neq {\cal O}'\,,
\label{rhocond1}\end{eqnarray}
where $A' = {\cal I} A $, with ${\cal I}$ given by Eq.~(\ref{aiso}).
We show that one can not extend a generic state $\rho_{(U,{\cal O})}$
to a collection of states on all neighborhoods satisfying (i)-(iii).\footnote{
We will need to generalize (\ref{rhocond1}) for this.}

The difficulty is associated with pairs of neighborhoods whose union is
time nonorientable.  Consider such a pair $(V,{\cal O}),(V',{\cal
O}')\in {\cal C}$.   Because $V\bigcup V'$ is time nonorientable, the
intersection $V\bigcap V'$ includes disjoint regions $\hat U $ and
$\check U$, such that the inherited time-orientations agree on $\hat U$
and disagree on $\check U$.  The restrictions of the states to $\hat U
\sqcup \check U$ are required to yield pairs of physically equivalent
states, seen by observers whose orientations agree on $\hat U$ and
disagree on $\check U$ (see (\ref{rhop2}) and (\ref{rhop3}) below for
the precise definition).  Without loss of generality, we may (by
choosing open subsets of $\hat U $ and $\check U$, if necessary) assume
that $\hat U$ and $\check U$ are globally hyperbolic and causally
convex.

Each choice of orientation ${\cal O}$ and ${\cal O}'$ gives a
well-defined quantum field theory on the {\it globally hyperbolic} spacetime
$U := \hat U\sqcup\check U, g|_U$.  The difficulty arises from
the relation between the two field theories, the requirement that for
each state $\rho$, there exist a physically equivalent state $\rho'$.
That is, corresponding to each orientation of $U$ is an algebra, ${\cal A}$ or
${\cal A}'$, generated by commuting subalgebras on $\hat U $ and
$\check U$; and for each plf $\rho$ on the algebra ${\cal A}'$ there
must be a plf $\rho'$ on ${\cal A}'$ whose values on observable
quantities agree with those of $\rho$ on the observables in ${\cal A}$
having the same physical meaning.

We assume that any symmetric element of ${\cal A}$ (or ${\cal A}'$) is an
observable.  Let $\hat{\cal A} := {\cal A}_{(\hat U, {\cal O}|{\hat
U})}$, $\hat{\cal A}' := {\cal A}'_{(\hat U, {\cal O}'|{\hat U})}$, and
denote by $\hat\phi(F)$ and $\hat\phi'(F)$ the smeared field operators
that generate $\hat{\cal A}$ and $\hat{\cal A}'$.  Then, because the
orientations agree, physically equivalent observables are related by
the isomorphism $\hat {\cal I}$ given by
\beq
\hat\phi'(F)= \hat {\cal I}\hat\phi(F),\qquad i = \hat{\cal I} i.
\eeq
Similarly, denote by $\check\phi(F)$ and  $\check\phi'(F)$ the smeared
fields generating $\check{\cal A} := {\cal A}_{(\check U, {\cal
O}|{\check U})}$ and $\check{\cal A}' := {\cal A}_{(\check U, {\cal
O}'|{\check U})}$.
Because the orientations ${\cal O}$ and ${\cal O}'$
disagree on $\check U$, physically equivalent observables in
$\check{\cal A}$ and $\check{\cal A}'$ are related by the antilinear
map $\check {\cal I}$, with
\beq
\check\phi'(F)= \check {\cal I}\check\phi(F),\qquad -i =
\check{\cal I} i.
\eeq

For any plf $\rho$ on ${\cal A}$, the corresponding plf $\rho'$ on
${\cal A}'$ must satisfy
\beq
\rho'(\hat {\cal I}\hat A)
= \rho(\hat A), \qquad \rho'(\check {\cal I}\check A) =
\rho(\check A),
\label{rhop}\eeq
for all symmetric $\hat A\in\hat {\cal A}$, $\hat A\in\check {\cal A}$.
Any element of $\hat {\cal A}$ (or of $\check {\cal A}$) can be
written as a linear combination, $\hat A = \hat A_1 +i\hat A_2$, of
symmetric elements, $A_1 = {1\over 2}(\hat A + \hat A^\dagger)$,
$A_2 = {1\over 2i}(\hat A - \hat A^\dagger)$.
Linearity of $\rho$ and $\rho'$ and Eq.~(\ref{rhop}) then imply
\begin{eqnarray}
\rho'(\hat {\cal I}\hat A) &= &\rho(A), \forall \hat A\in \hat {\cal A}
\nonumber\\
\rho'[(\check {\cal I}\check A)^\dagger] &= &\rho(A),
\forall \check A\in \check {\cal A}\,.
\label{rhop1}\end{eqnarray}

Now the product $\hat A\check A$ of two symmetric elements $\hat
A\in\hat {\cal A}$, $\hat A\in\check {\cal A}$ is again symmetric and
corresponds to the observable $\hat {\cal I}\hat A\ \check {\cal
I}\check A \in {\cal A}'$.  Thus, we require
\beq
\rho'(\hat {\cal I}\hat A\ \check {\cal I}\check A) = \rho(\hat A \check A).
\label{rhop2}\eeq
This final requirement uniquely determines $\rho'$, because any element
of $\hat {\cal A}$ can be written as a linear combination of terms of
the form $\hat A\check A$ with $\hat A$ and $\check A$ symmetric. The
resulting $\rho'$ satisfies
\beq
\rho'[\hat {\cal I}\hat A\ (\check {\cal I}\check A)^\dagger] =
\rho(\hat A \check A),
\forall \hat A\in \hat {\cal A},\ \check A\in \check {\cal A}.
\label{rhop3}\eeq
(This equation makes sense, because the factorization is unique up to a
complex number $c$ -- one can represent the same algebra element as the
product $(c\hat A)(c^{-1}\check A)$
-- and the maps
$\hat A\mapsto \hat{\cal I}\hat A$ and
$\check A\mapsto (\check{\cal I}\check A)^\dagger$ are both linear.)

We claim that the linear function $\rho'$ defined by Eqs.~(\ref{rhop1})
and (\ref{rhop2}) is not in general positive.  Let $\hat{A} \in
\hat{{\cal A}}$, $\check{B} \in \check{{\cal A}}$ and a state
$\rho_0$ satisfy
\begin{equation}
\left[ \hat{A},\hat{A}^{\dagger} \right]  =  1,\;\;\;\;
\left[ \check{B},\check{B}^{\dagger} \right]  =  1\,,
\end{equation}
and
\begin{equation}
\rho_0 ( \hat{A}^{\dagger}\hat{A} ) =
\rho_0 ( \check{B}^{\dagger}\check{B} ) = 0\,.  \label{at1}
\end{equation}
Using the Schwarz inequality,
$| \rho_0 ( X\hat{A})|^2 \leq
\rho_0 ( X X^{\dagger})\rho_0 ( \hat{A}^{\dagger}\hat{A})$,
for any operator $X$,
one finds $\rho_0 ( X \hat{A} ) = 0$, and similarly,
$\rho_0 ( X \check{B} ) = 0$.
For the special case where
$\hat{A} = \hat{\phi}(F)$ and $\check{B} = \check{\phi}(G)$ for some
(complex) functions $F$ and $G$ with support in $\hat{U}$ and $\check{U}$,
respectively, and where $\rho_0 = |\psi\rangle\langle\psi |$ for a state
$|\psi\rangle$ in a Hilbert space, assumption (\ref{at1}) implies
$\hat{A}|\psi\rangle = \check{B}|\psi\rangle = 0$.

Now define a plf $\rho_1$ by
\begin{equation}
\rho_1(X) := \frac{1}{1 + c^2}\rho_0 [
(1+c\hat{A}\check{B})X(1+c\hat{A}^{\dagger}\check{B}^{\dagger})]\,,
\end{equation}
where $c > 0$.  We
will show that the corresponding
state $\rho'_1$ satisfying
(\ref{rhop3})
is not positive.
Consider the positive operator ${\cal I}O \in {\cal A}'$ defined by
\begin{equation}
{\cal I}O := ((\hat{{\cal I}}\hat{A})^{\dagger} - (\check{{\cal I}}
\check{B})^{\dagger})
(\hat{{\cal I}}\hat{A} - \check{{\cal I}}\check{B})\,.
\end{equation}
According to
(\ref{rhop3}),
we have
\begin{eqnarray}
\rho'_1({\cal I}O) & = & \rho_1(\hat{A}^{\dagger}\hat{A})
+ \rho_1(\check{B}^{\dagger}\check{B})
- 2{\rm Re}[\rho_1(\hat{A}\check{B})] \nonumber \\
 & = & \frac{2c(c-1)}{1+c^2}\,.
\end{eqnarray}
Thus, $\rho'_1$ is nonpositive if $c < 1$.

The state $\rho_0$ satisfying (\ref{at1}) is not likely to be physically
realistic, considering the fact that no
annihilation operator for the vacuum state in Minkowski space can be
localized.
However, by choosing $\hat{A}$ and $\check{B}$ to be approximate
annihilation operators
for high-frequency modes, one should be able to satisfy
condition (\ref{at1}) approximately
for a physically realistic state.
Moreover, one can show a similar nonpositivity with a weaker and
physically realistic condition
\begin{equation}
\rho [ (\hat{A}^{\dagger}\hat{A})^3 ],\
\rho [ (\check{B}^{\dagger}\check{B})^3 ] < \frac{1}{\sqrt{24}}\,,
\label{at2}
\end{equation}
as demonstrated in Appendix A.

In fact it is likely that a linear function in a large globally
hyperbolic neighborhood $U_L$ which contains an initial surface
$\Sigma$ except for a measure-zero boundary cannot be positive under
the assumption of reasonable short-distance behavior.  The argument is
as follows.  Consider a small neighborhood $U_S$ which contains part of
the above-mentioned boundary. The set $U_S\bigcap U_L$ can be
approximated near the measure-zero boundary by the left and right
Rindler wedges in Minkowski space.  Now, any physically realistic state
on ${\cal A}_{(U_S,{\cal O}|U_S)}$ should behave like the Minkowski
vacuum for high-frequency modes.  The Minkowski vacuum can be expressed
as a linear combination of products of left and right Rindler states
\cite{unruh}.  On the other hand, in $U_L$ the approximate left and
right Rindler wedges have opposite time directions.  Then, a
construction similar to that given above, with the left and right
Rindler wedges playing the role of $\hat{U}$ and $\check{U}$, is likely
to show that there is a nonpositive linear function on
${\cal A}_{(U_L,{\cal O}|U_L)}$
whose restriction to $U_S\bigcap U_L$ corresponds to a plf on
${\cal A}_{(U_S,{\cal O}|U_S)}$.

\subsection{Collections of local states associated with restricted atlases}
We have argued that one cannot consistently define a collection of
local states $\rho_{(U,{\cal O})}$ on an atlas that includes pairs of
neighborhoods whose union is time nonorientable.  We now consider
restricted atlases of neighborhoods satisfying (i)-(iii) together with
the additional condition:\\ (iv) The closure of $U\bigcup V$ is
time-orientable for each $U,V\in C$.  One maximizes the information
available in such a collection of states by considering a maximal
collection $C$ of neighborhoods (i)-(iv) and covering $M$.  Again we
denote by ${\cal C}$ the corresponding collection of all oriented
neighborhoods, ${\cal C} =\{(U,{\cal O}) | U\in C\}$.

Condition (iv) is necessary to extend a local state to a collection of
states on neighborhoods of the atlas ${\cal C}$.  We now show that it
is sufficient.  We first consider the problem in the globally hyperbolic
context and then return to our time-nonorientable spacetimes.

In a globally hyperbolic spacetime, one is free to choose a state
$\rho_U$ on the algebra of any globally hyperbolic subspacetime $U,
g|_U$ that satisfies conditions (i)-(iii).  One can then construct a
global state of which $\rho_U$ is the restriction to $U$. Elements of
the *-algebra ${\cal A}_{(U,{\cal O}_U)}$ are linear combinations of
products of
Klein-Gordon inner products $\omega_U(\hat\phi, f)$ that involve only
data for $f$ on a Cauchy surface $\Sigma_U$.  A state is specified by
the expectation values, $\rho_U[\omega_U(\hat\phi,
f)\cdots\omega_U(\hat\phi, g)]$, or, equivalently, by $n$-point
distributions,
$\rho(\phi(x)\cdots\phi(y)),\ \rho(\phi(x)\cdots\phi(y)\pi(z)),\dots ,
\rho(\pi(x)\cdots\pi(y))$,
where $x, y, \ldots , z \in \Sigma_U$.
  A state $\rho_U$ can be extended to a state
on the larger spacetime by adjoining values of $\rho[\omega(\hat\phi,
f) \dots \omega_U(\hat\phi, g)]$, where at least one of $f, \dots, g$
has support on the part of the full Cauchy surface $\Sigma$ that lies
outside of $U$.

The simplest way to extend $\rho_U$ to a global state is as follows.
First, one specifies a state in the interior of the domain of
dependence of $\Sigma\backslash \Sigma_U$.  Call this state
$\rho_{\bar{U}}$.  Then, we define the global state $\rho := \rho_U
\otimes \rho_{\bar{U}}$.  That is,
\begin{eqnarray}
& & \rho\left[ \omega(\hat{\phi},f_1)\cdots\omega(\hat{\phi},f_n)
\omega(\hat{\phi},g_1)\cdots\omega(\hat{\phi},g_m)\right] \nonumber \\
& & :=
\rho_U\left[ \omega_U(\hat{\phi},f_1)\cdots\omega_U(\hat{\phi},f_n)\right]
\rho_{\bar{U}}\left[ \omega_{\bar{U}}(\hat{\phi},g_1)\cdots
\omega_{\bar{U}}(\hat{\phi},g_m)\right]\,,
\end{eqnarray}
where $f_1,\ldots , f_n$ involve only data on $\Sigma_U$, whereas
$g_1,\ldots,g_m$ involve only data on $\Sigma\backslash\Sigma_U$.  The
function $\rho$ is positive if $\rho_U$ and $\rho_{\bar{U}}$ are.

The global state $\rho$ is rather unphysical in the sense that there is
no correlation between the field operators on $\Sigma_U$ and those on
$\Sigma\backslash\Sigma_U$.  It is also likely that the (renormalized)
stress-energy tensor will be singular on the light cone of the boundary
of $\Sigma_U$, since the state $\rho$ is analogous to the direct
product of the left and right Rindler vacua in Minkowski space, which
is known to possess such singularities \cite{cd}.  It will be
interesting to establish ``extendibility" of states under some physical
requirements, such as the absence of singularity in the stress-energy
tensor.  However, not much is known about these issues, as far as we
are aware.  We suspect, but have not verified, that if one restricts
consideration to Hadamard states, then one can extend such a state on
$U$ with suitable behavior at $\partial U$ to a Hadamard state on the
full spacetime.

Now, consider a Lorentzian universe-from-nothing, a time-nonorientable
spacetime $M, g$ of the form (\ref{lor}),
\begin{equation}
M = (\sRR\times \tilde\Sigma)/Q\,,
\label{mfld}
\end{equation}
with
\begin{equation}
Q = T\times I\,,
\end{equation}
where $T(t) = -t$, for $t\in\sRR$, and $I$ is an involution of
$\tilde\Sigma$ with no fixed points.  We will denote the orientable
double cover by $p:\tilde  M\rightarrow M$, where $\tilde  M=
\sRR\times \tilde \Sigma$.  We will write $\Sigma = \tilde \Sigma/I$,
and, for simplicity, we will denote by $\tilde  \Sigma$ and $\Sigma$ the
particular copies $\tilde  \Sigma\times\{0\}$ and $p(\tilde
\Sigma\times\{0\})$ of these 3-manifolds in $\tilde  M$ and $M$.  The
metric $g$ is chosen to make $\tilde  M,\ \tilde  g = p_* g$ globally
hyperbolic, with Cauchy surface $\tilde \Sigma$.  Then $\Sigma$ is
an initial value surface of $M, g$.

Let ${\cal C}$ be an atlas for $M$ satisfying conditions (i)-(iv).  On
any oriented neighborhood $(U, {\cal O})$ in the atlas, one can freely
specify a state $\rho_{(U,\,{\cal O})}$.  We are to extend the state to
a collection of states on the algebras associated with ${\cal C}$,
satisfying Eq.~(\ref{rhocond1}).  To do this, we first lift the local
state to the globally hyperbolic covering spacetime $\tilde  M, \tilde
g$, extend that lifted state to an antipodally symmetric state on
$\tilde  M$, and then use the antipodally symmetric state on $\tilde  M$
to provide a collection of states on the atlas ${\cal C}$.

Given an orientation $\tilde {\cal O}_{\tilde  M}$ on $\tilde  M$, there
is a 1-1 correspondence between oriented neighborhoods $(U,{\cal O}_U)$
and neighborhoods $\tilde  U$ of $\tilde  M$.  (Because the two lifts of
$(U,{\cal O}_U)$ to $\tilde  M$ have opposite time orientation,  there
is a unique lift of $(U,{\cal O})$ to an oriented neighborhood $(\tilde
U, \tilde  {\cal O}_U)$ for which the orientations induced by ${\cal
O}_U$ and $\tilde {\cal O}_{\tilde  M}$ agree.) The atlas ${\cal C}$ on
$M$ is thus mapped to an atlas $\tilde {\cal C}$ of oriented
neighborhoods that cover $\bar M$, all with orientation
$\tilde {\cal O}_{\tilde  M}$.

The isomorphisms $p|_{\tilde  U}: \tilde  U\rightarrow U$ induce
algebra isomorphisms
$\tilde {\cal A}_{(\tilde  U,\tilde  {\cal O})}
\rightarrow {\cal A}_{(U, {\cal O})}$,
given by
\begin{equation}
\phi(F)\rightarrow\tilde \phi(F\circ p|_{\tilde  U})\,.
\end{equation}
Thus the family of algebras (including both orientations) associated
with the atlas ${\cal C}$ of $M$ is identical to the family of algebras
(all with the same orientation) associated with the oriented atlas
$\tilde {\cal C}$ of $\tilde M$.  Because $\tilde M$ is globally
hyperbolic, one can regard all local algebras as subalgebras of a
global algebra $\tilde {\cal A}_{\tilde  M}$ associated with the
orientation $\tilde {\cal O}_M$.

The family of algebras on $M$, however, has an additional structure
that plays no role in quantum field theory on the covering
space itself, namely the collection of antilinear isomorphisms between
oppositely oriented neighborhoods $(U,{\cal O})$ and $(U,\check {\cal
O})$.  On $\tilde M$, these can be regarded as antilinear isomorphisms
between the algebras associated with antipodally related
neighborhoods, or, equivalently, as an antilinear isomorphism ${\cal Q}$
of the global algebra  $\tilde {\cal A}_{\tilde  M}$ given by
\begin{eqnarray}
{\cal Q}\tilde \phi(F)&=&\tilde \phi (F\circ Q), \nonumber\\
{\cal Q} i &=& - i\,. \label{aisoc}
\end{eqnarray}

A collection of states associated with an atlas ${\cal C}$ of $M, g$
that satisfies the overlap conditions (\ref{rhocond1}) can then be
regarded as a collection of states associated with the atlas
$\tilde {\cal C}$ on $\tilde  M$; it must satisfy the usual overlap
condition for a collection of states on the oriented atlas
$\tilde {\cal C}$ and the additional condition
\begin{equation}
\rho_{\tilde  U}(\tilde  A) = \rho_{Q(\tilde  U)}({\cal Q} \tilde A)\,.
\label{rhocondc}
\end{equation}
If the collection of states corresponds to a single state on
$\tilde {\cal A}_{\tilde  M}$, the additional condition is simply the
statement that it is {\em antipodally symmetric}.

Given a plf $\rho_U$ on the algebra ${\cal A}_{(U,{\cal O}_U)}$ associated
with an oriented neighborhood $(U,{\cal O}_U)\in {\cal C}$, one
can extend it as follows to a collection of states on all algebras
associated with  ${\cal C}$.  Again denote by $\Sigma$ an initial
value surface of $M$ shared by $U$ -- for which $\Sigma_U$ is a
Cauchy surface for $U, g|_U$ -- and denote by $\tilde  \Sigma$
the corresponding
Cauchy surface of $\tilde  M$.  Then $\rho_U$ can be regarded as
a plf $\rho_{\tilde  U}$ on $\tilde  {\cal A}_{\tilde  U}$; and
$\rho_{Q(\tilde  U)}$ given by Eq.~(\ref{rhocondc}) can be regarded
as a plf on the disjoint neighborhood $Q(\tilde  U)$.  We can now use
the construction given above for globally hyperbolic spacetimes
to extend it to a plf $\tilde \rho_0$ on $\tilde  {\cal A}_{\tilde  M}$.
The plf $\tilde \rho_0$ will not in general
be antipodally symmetric, but we can obtain
a plf that is both antipodally symmetric and positive by writing
\begin{equation}
\tilde \rho = {1\over2}(\tilde \rho_0 + \tilde \rho_0\circ{\cal Q})\,.
\end{equation}
Then the restrictions of $\tilde \rho$ to the subalgebras $\hat{\cal
A}_{\tilde  U}$ (and the identification of the subalgebras with
algebras associated with ${\cal C}$) yield a collection of states
$\rho_{(U,{\cal O})}$ satisfying the overlap conditions
(\ref{rhocond1}) as required.

What are the implications of condition (iv), restricting possible
atlases to neighborhoods small enough that no two neighborhoods contain
an orientation-reversing curve?  If our universe has the topology of
antipodally identified deSitter space and a volume larger than the
currently visible universe, one can choose an atlas that includes
open sets with spatial extent as large as the visible universe.  This
is enough to allow one mechanically to replicate the observable part of
quantum field theory with a collection of states and algebras
associated with an atlas restricted by condition (iv).

{}From a more fundamental point of view, however, the theory is not
satisfactory.  Let us reiterate, in hindsight, the objections mentioned
earlier.  Because the correlations that are allowed depend on the
atlas, one obtains a different theory for every choice of atlas.  There
is no unique way to pick a largest atlas satisfying conditions
(i)-(iv), and thus no unique theory.  The missing correlations mean
that the information contained in a collection of states associated
with neighborhoods that that cover an initial value surface of $M$ is
incomplete.\footnote
{One can extend a collection of states given on an atlas ${\cal C}$ in
more than one way to an antipodally symmetric state on $\tilde M$,
because there are sets of points (and small neighborhoods about them)
among which no correlations are defined.  Here is an example of two
antipodally symmetric algebraic states on deSitter space that are
extensions of the same collection of states on an atlas ${\cal C}$.
Let $f$ and $g$ be complex solutions to the scalar wave equation on $M$
(antipodally-identified deSitter), whose initial data on an initial
value surface $\Sigma$ have support on disjoint neighborhoods $U$ and
$V$ of $\Sigma$ that do not both belong to a single neighborhood in
${\cal C}$.  Then no correlations between points of $U$ and $V$ are
defined by the collection of states on $M$.
Let also $i\omega(\bar{f},f),\ i\omega(\bar{g},g) > 0$ with respect to a
chosen time direction.
Then let  $\tilde f$ and $\tilde g$ be the (normalized)
antipodally symmetric lifts of $f$ and $g$ to $\tilde M$.
(Thus, $\tilde{f}$ and $\tilde{g}$ are mapped to their complex conjugates
under the antipodal map.)
 Pick as a Hilbert space ${\cal H}$ on $\tilde M$ a Fock space
associated with a state $|0\rangle$ that is annihilated by the
annihilation operators $\tilde\omega(\tilde{\phi}, \tilde f)$ and
$\tilde\omega(\tilde{\phi},\tilde g)$
corresponding to $f$ and $g$.  Define an antipodally
symmetric state by
$\vert\psi\rangle\ =
\tilde\omega(\tilde\phi,\tilde f)^{\dagger}
\tilde\omega(\tilde\phi,\tilde g)^{\dagger}|0\rangle$.
Let $\rho = 1/2(|0\rangle\langle0|+|\psi\rangle\langle\psi|)$.
Then, for small $e$,
$\rho' = 1/2(|0\rangle\langle0|+|\psi\rangle\langle\psi|)+e
(|0\rangle\langle\psi| + |\psi\rangle\langle0|)$
is positive, and $\rho$ and $\rho'$ give the same family of states on $M$.}
One is not entitled to regard a state as an assignment of local
states to a collection of local algebras restricted by condition (iv),
because a ``state'' so defined has no unique time-evolution.  Finally,
in order to extend a local state to a family of states on an atlas
${\cal C}$, we used an antipodally symmetric state on $\tilde M$.  This
suggests that by artificially restricting the collection of
neighborhoods, we are simply providing a way to interpret an
antipodally symmetric state: one can make such a state consistent if
one chooses neighborhoods small enough.  But if one includes all
globally hyperbolic neighborhoods that inherit their causal structure
from the spacetime $M, g$, an antipodally symmetric state on the
covering space does not yield a consistent collection of local states
on $M, g$.

A ``theory'' in which the correlations that one can measure are limited
and depend on the choice of atlas might be more acceptable if one
regards, say, a path-integral formulation as fundamental and relegates
the usual quantum field theory to a subsidiary position. In a
path-integral interpretation where the measuring instrument is included
in the system, the measurements that can be made depend on the state of
the instrument.  Different states of the instrument will pick out
different observables.  Measuring a correlation between field operators
at two spacetime points plausibly requires a choice of time-orientation
at each of the two points in such a theory.  This suggests that each
state might carry with it an implicit atlas (or partial atlas) of
oriented neighborhoods, covering at least regions of spacetime where
measurements are effectively made.  But this is a much weaker structure
than the one we have considered, and it suggests that -- if there is to
be a sensible quantum field theory on time-nonorientable spacetimes --
condition (\ref{rhop3}), which presupposes a physical meaning of
correlations irrespective of time-orientations, may be too strong.

\section{A note on fermion fields}

It is common in the general-relativity literature to regard
two-component spinors as fields built from the fundamental
representation of $SL(2,\CC)$\cite{bichteler,geroch1,pr,wald}.  Chiral
fermions, however, are really acted on by a larger group that includes
time reversal.  That is, a Weyl spinor belongs to an action of a
covering group of that subgroup $L_+$ of the full Lorentz group
comprising the component of the identity and the component of
time-reversing, space-preserving Lorentz transformations.  Readers of
earlier work\cite{pr,gh1,gh2} may have been left with the misimpression
that one cannot define two-component spinors on time-nonorientable
spacetimes, and the present section, summarizing work from \cite{fh2},
(see also the sequel \cite{gc} by Chamblin and Gibbons), is intended as
a clarification.

The group $L_+$ has two double covers, Sin$^+$ and Sin$^-$, depending
on the sign of ${\cal T}^2$, where ${\cal T}$ is either of the two
elements of the covering group that correspond to the choice $T$ of
time-reversal.   Only Sin$^-$ acts on the usual two-component spinors
associated with Weyl neutrinos in Minkowski space.  In an orientable
spacetime, the difference between an $SL(2,\CC)$-spinor structure and a
Sin$^-$-spinor structure is unimportant, unless one wishes explicitly
to discuss time reversal.  In a time-nonorientable spacetime, however,
the difference is essential.  Because one cannot pick a bundle of
time-oriented frames, time-nonorientable spacetimes have no
$SL(2,\CC)$-spinor structure, and two-component spinors rely for their
definition on an action of a the covering group Sin$^-$ of $L_+$.
Other authors have considered generalized spinor structures on generic
nonorientable spacetimes
\cite{karoubi,whiston,dabrowski,chamblin,gibbons}.  For these generic
spacetimes, the situation is somewhat different, because one must
consider actions of the full Lorentz group; and the usual action of
parity requires four-component spinors.

Every Lorentzian universe-from-nothing, every spacetime of the form
(\ref{lor}), has a Sin$^+$-spinor structure, but only a subclass has a
Sin$^-$-spinor structure.  Inequivalent Sin$^+$- and Sin$^-$-spinor
structures correspond to members of two classes of homomorphisms from
$\pi_1(\tilde M)$ to $\ZZ_2$, where $\tilde M$ is the orientable double
cover of the spacetime manifold $M$.

A precise statement is as follows:\\
{\sl Proposition}.  Let $M, g$ be a spacetime of the form (\ref{lor})
and let $\tilde M$ be its orientable double cover.  Then the inequivalent
Sin$^+$-spinor structures (Sin$^-$-spinor structures) are in 1-1
correspondence with homomorphisms $h\in H^1(\tilde M)$ that respect
the antipodal map $A$ and for which $h[\widetilde{c^2}] = +1$
($h[\widetilde{c^2}] = -1$), for every time reversing curve $c$ in $M$.
In particular, every such spacetime has a Sin$^+$-spinor structure.

Here $\widetilde{c^2}$ is a lift of $c^2$ to $\tilde M$.  A field of
two-component spinors is then a cross section of a bundle associated to
a Sin$^-$-spinor structure.  Lorentzian universes-from-nothing for
which $\Sigma$ is a 3-torus have both Sin$^+$ and Sin$^-$-spinor
structures, while antipodally identified deSitter space has only a
Sin$^+$-spinor structure and so does not admit global fields of the
usual kind of chiral fermions.  Even in a time-nonorientable spacetime
that allows the usual 2-component spinors, however, one cannot
construct a global Lagrangian density that violates time-reversal
invariance.

\acknowledgments

We thank Bernard Kay for lengthy and helpful conversations, Robert Wald
for tutoring us in algebraic field theory, and Chris Isham, Gary
Gibbons and Chris Fewster for useful discussions.

This research was supported in part by NSF Grants PHY91-05935 and
PHY92-20644 and by Schweizerischer Nationalfonds.

\appendix
\section{Nonpositivity of $\rho'_1$ with condition (3.15)}
In this appendix we prove that $\rho'_1$ defined in
Sect.~\ref{sec:noor}, with a slight technical
modification, is nonpositive under condition (\ref{at2}).
Let us first prove some general inequalities.  By noting
\begin{equation}
\rho_0 [ (\hat{A}^{\dagger}\hat{A})^2 ]
= \rho_0 [\hat{A}^{\dagger}\hat{A}]
+ \rho_0 [ (\hat{A}^{\dagger})^2 (\hat{A})^2 ]
\end{equation}
and
\begin{equation}
\rho_0 [ (\hat{A}^{\dagger}\hat{A})^3 ] =
\rho_0 [ (\hat{A}^{\dagger}\hat{A})^2 ] +
\rho_0 [ (\hat{A}^{\dagger})^2 \hat{A}^2 ]
+ \rho_0[ (\hat{A}^{\dagger})^2 \hat{A} \hat{A}^{\dagger}\hat{A}^2 ]\,,
\end{equation}
we obtain
\begin{equation}
\rho_0[\hat{A}^{\dagger}\hat{A} ] \leq
\rho_0[ (\hat{A}^{\dagger}\hat{A})^2 ] \leq
\rho_0[ (\hat{A}^{\dagger}\hat{A})^3 ] < \epsilon\,,  \label{at4}
\end{equation}
where $0 < \epsilon < 1/\sqrt{24}$.
Using these inequalities, we find
\begin{equation}
\rho_0[(\hat{A}\hat{A}^{\dagger})^3]
= \rho_0[(\hat{A}^{\dagger}\hat{A} + 1)^3]
\leq 1 + 7\epsilon\,.
\end{equation}
Then, using the Schwarz inequality, we have
\begin{equation}
|\rho_0[\hat{A}^{\dagger}\hat{A}\check{B}^{\dagger}\check{B}]|^2
\leq \rho_0[(\hat{A}^{\dagger}\hat{A})^2]
\rho_0[(\check{B}^{\dagger}\check{B})^2]\,.
\end{equation}
Hence
\begin{equation}
\rho_0[\hat{A}^{\dagger}\hat{A}\check{B}^{\dagger}\check{B}]
\leq \epsilon\,.
\label{at5}
\end{equation}
Next we will prove
\begin{equation}
\rho_0[(\hat{A}\hat{A}^{\dagger})^2\check{B}\check{B}^{\dagger}]
+ \rho_0[\hat{A}\hat{A}^{\dagger}(\check{B}\check{B}^{\dagger})^2]
\leq 2(1 + 7\epsilon)\,.  \label{at3}
\end{equation}
Define
\begin{equation}
s_1 := \rho_0[ (\hat{A}\hat{A}^{\dagger})^2\check{B}\check{B}^{\dagger}]
\end{equation}
and
\begin{equation}
s_2 := \rho_0[ \hat{A}\hat{A}^{\dagger}(\check{B}\check{B}^{\dagger})^2]\,.
\end{equation}
By using the Schwarz inequality
$|\rho_0(X^{\dagger}Y)|^2 \leq \rho_0(X^{\dagger}X)\rho_0(Y^{\dagger}Y)$
with
$X = \hat{A}^{\dagger}\hat{A}\hat{A}^{\dagger}$ and
$Y = \hat{A}^{\dagger}\check{B}\check{B}^{\dagger}$,
we find $s_1^2 \leq s_0 s_2$, where $s_0 = 1 + 7\epsilon$.  In a similar
manner we find $s_2^2 \leq s_0 s_1$.  Then, from these
two inequalities we have $s_1 s_2 \leq s_0^2$.  Hence
\begin{equation}
(s_1 + s_2)^2 \leq s_0 (s_1 + s_2) + 2s_0^2\,.
\end{equation}
{}From this we immediately obtain (\ref{at3}), i.e.,
$s_1 + s_2 - 2s_0 \leq 0$.

Now, given a plf $\rho_0$ satisfying (\ref{at2}), we define a new plf
$\rho_1$ by
\begin{equation}
\rho_1(X) := k\rho_0[(1 + ce^{i\alpha}\hat{A}\check{B})X
(1 + ce^{-i\alpha}\hat{A}^{\dagger}\check{B}^{\dagger})]\,,
\end{equation}
where $k$ is the normalization factor given by
\begin{eqnarray}
k^{-1} & = & \rho_0\left[ (1+ce^{i\alpha}\hat{A}\check{B})
(1+ce^{-i\alpha}\hat{A}^{\dagger}\check{B}^{\dagger})\right] \nonumber \\
 & = &
\rho_0\left[ (1+ce^{-i\alpha}\hat{A}^{\dagger}\check{B}^{\dagger})
(1+ce^{i\alpha}\hat{A}\check{B})\right] + c^2\rho_0\left( \left[
\hat{A}\check{B},\hat{A}^{\dagger}\check{B}^{\dagger}
\right]\right)\,.
\end{eqnarray}
The last term equals
$c^2(\hat{A}^{\dagger}\hat{A} + \check{B}^{\dagger}\check{B} + 1)$.  Hence
$k^{-1} > 0$ and $\rho_1$ is indeed a plf.
Next we consider a positive operator ${\cal I}O$ in the other neighborhood
$V'$ defined by
\begin{equation}
{\cal I}O := ((\hat{{\cal I}}\hat{A})^{\dagger} -
e^{i\alpha}(\check{{\cal I}}
\check{B})^{\dagger})(\hat{{\cal I}}\hat{A}
- e^{-i\alpha}\check{{\cal I}}\check{B})\,.
\end{equation}
Then, according to (\ref{rhop2}),
the operator ${\cal I}O$ takes the following value
for the state $\rho'_1$:
\begin{equation}
\rho'_1({\cal I}O) = \rho_1(\hat{A}^{\dagger}\hat{A})
+ \rho_1(\check{B}^{\dagger}\check{B})
- 2{\rm Re}\left[e^{i\alpha}\rho_1(\hat{A}\check{B})\right]\,.
\end{equation}
Hence we have
\begin{eqnarray}
k^{-1}\rho'_1({\cal I}O) & = & \rho_0\left[
(1+ce^{i\alpha}\hat{A}\check{B})(\hat{A}^{\dagger}\hat{A}
+\check{B}^{\dagger}\check{B})
(1+ce^{-i\alpha}\hat{A}^{\dagger}\check{B}^{\dagger})\right] \nonumber \\
&  & -2{\rm Re}\left( e^{i\alpha}\rho_0
\left[ ( 1+ce^{i\alpha}\hat{A}\check{B})
\hat{A}\check{B}(1+ce^{-i\alpha}\hat{A}^{\dagger}\check{B}^{\dagger})\right]
\right)\,.
\end{eqnarray}
When we expand this expression, the sum of the terms proportional to
$e^{\pm i\alpha}$ or  $e^{\pm 2i\alpha}$ takes the form
$$
A\cos(\alpha+\delta_1) + B\cos(2\alpha + \delta_2).
$$
This can be made nonpositive by choosing $\alpha$ appropriately.
Hence we may drop these terms and obtain
\begin{eqnarray}
k^{-1}\rho'_1({\cal I}O) & \leq &
\rho_0(\hat{A}^{\dagger}\hat{A} + \check{B}^{\dagger}\check{B})
-2c\rho_0(\hat{A}\hat{A}^{\dagger}\check{B}\check{B}^{\dagger}) \nonumber \\
& & + c^{2}\left(\rho_0\left[
(\hat{A}\hat{A}^{\dagger})^2\check{B}\check{B}^{\dagger}\right]
+ \rho_0\left[
\hat{A}\hat{A}^{\dagger}(\check{B}\check{B}^{\dagger})^{2}\right]\right)\,.
\end{eqnarray}
By using
\begin{equation}
\rho_0(\hat{A}\hat{A}^{\dagger}\check{B}\check{B}^{\dagger})
= \rho_0[(\hat{A}^{\dagger}\hat{A}+1)(\check{B}^{\dagger}\check{B}+1)]\,,
\end{equation}
we find
\begin{eqnarray}
k^{-1}\rho_1({\cal I}O) & \leq &
(1-2c)\left[ \rho_0(\hat{A}^{\dagger}\hat{A})
+ \rho_0(\check{B}^{\dagger}\check{B})\right]
-2c\left[ 1+ \rho_0
(\hat{A}^{\dagger}\hat{A}\check{B}^{\dagger}\check{B})\right]
 \nonumber \\
& & + c^{2}\left(\rho_0\left[
(\hat{A}\hat{A}^{\dagger})^2\check{B}\check{B}^{\dagger}\right]
+ \rho_0\left[
\hat{A}\hat{A}^{\dagger}(\check{B}\check{B}^{\dagger})^{2}\right]\right)\,.
\end{eqnarray}
Using inequalities (\ref{at4}), (\ref{at5}) and (\ref{at3}) and assuming
$c < 1/2$, we obtain
\begin{equation}
k^{-1}\rho'_1({\cal I}O) \leq 2c^2 (1+7\epsilon) - 2c(1+2\epsilon) +
2\epsilon\,.
\end{equation}
For the right-hand side to have a
negative value for some $c$, it is sufficient to have
\begin{equation}
(1+2\epsilon)^2 - 4(1+7\epsilon)\epsilon > 0\,.
\end{equation}
Hence, if
\begin{equation}
\epsilon < \frac{1}{\sqrt{24}}
\end{equation}
and
\begin{equation}
\frac{1+2\epsilon -\sqrt{1-24\epsilon^2}}{2(1+7\epsilon)} < c<\frac{1}{2}\,,
\end{equation}
then $\rho'_1({\cal I}O) < 0$ as claimed.

\newpage
\begin{figure}
\caption{ Diametrically opposite points of the past boundary $\check\Sigma$
are identified to construct a smooth Lorentzian universe with no past
boundary.} \label{f1}
\end{figure}
\begin{figure}
\caption{For a 2-dimensional cylinder, the resulting spacetime is a
M\"obius strip whose median circle $\Sigma$ is obtained by the
identification of points in $\check\Sigma$.} \label{f2}
\end{figure}
\end{document}